\documentclass{PoS}

\newcommand{\mc}[3]{\multicolumn{#1}{#2}{#3}}

\title{Optical polarisation variability of narrow line Seyfert 1 galaxies}

\ShortTitle{Optical polarisation variability of narrow line Seyfert 1 galaxies}

\author{\speaker{Emmanouil Angelakis}\\
         Max-Planck-Institut f\"ur Radioastronomie, Auf dem H\"ugel 69, Bonn 53121, Germany \\
        E-mail: \email{eangelakis@mpifr.de}}

\author{S. Kiehlmann\\
	Owens Valley Radio Observatory, California Institute of Technology, Pasadena, CA 91125, USA
		}

\author{I. Myserlis\\
         Max-Planck-Institut f\"ur Radioastronomie, Auf dem H\"ugel 69, Bonn 53121, Germany 
		}

\author{D. Blinov\\
        Department of Physics and Institute for Plasma Physics, University of Crete, 71003, Heraklion, Greece
		}

\author{J. Eggen\\
		Center for Research and Exploration in Space Science, NASA, Greenbelt, MD 20771, USA\\
		Department of Astronomy, University of Maryland, College Park, MD 20742, USA
		}

\author{R. Itoh\\
        Department of Physics, School of Science, Tokyo Institute of Technology, 2-12-1 Ohokayama, Meguro, Tokyo 152-8551, Japan
		}

\author{S. Komossa\\
         Max-Planck-Institut f\"ur Radioastronomie, Auf dem H\"ugel 69, Bonn 53121, Germany 
		}

\author{N. Marchilli\\
		Istituto di Astrofisica e Planetologia Spaziali Via Fosso del Cavaliere 100, 00133, Rome, Italy
		}

\author{J. A. Zensus\\
         Max-Planck-Institut f\"ur Radioastronomie, Auf dem H\"ugel 69, Bonn 53121, Germany 
		}

\abstract{We have monitored the $R$-band optical linear polarisation of ten jetted NLSy1 galaxies with the aim to quantify their variability and search for candidate long rotation of the polarisation plane. In all cases for which adequate datasets are available we observe significant variability of both the polarisation fraction and angle. In the best-sampled  cases we identify candidate long rotations of the polarisation plane. We present an approach that assesses the probability that the observed phenomenology is the  result of pure noise. We conclude that although  this possibility cannot be excluded it is much more likely that the EVPA undergoes an intrinsic evolution. We compute the most probable parameters of the intrinsic event which forecasts events consistent with the observations. In one case we find that the EVPA shows a preferred direction which, however, does not imply any dominance of a toroidal or poloidal component of the magnetic field at those scales.}

\FullConference{Revisiting narrow-line Seyfert 1 galaxies and their place in the Universe - NLS1 Padova\\
		9-13 April 2018 \\
		Padova Botanical Garden, Italy}

\begin{document}

\section{Introduction: why study Narrow line Seyfert 1 galaxies}
The term narrow-line Seyfert 1 galaxies (NLSy1s) labels the subset of active galactic nuclei (AGN) with narrow width of the broad Balmer emission line (FWHM(H$\beta$)$\le 2000$~km~s$^{-1}$), and weak forbidden lines with [O$_\mathrm{III}$]$\lambda$5007/H$\beta$ $<3$ \cite{1985ApJ..297...166,1989ApJ...342..224G,2006ApJS..166..128Z}. NLSy1s are thus associated with black hole masses in the range $10^{6}$--$10^{8}$~M$_{\odot}$ \cite{2006AJ....132..531K,2008ApJ...685..801Y,2012AJ....143...83X,2015A&A...575A..13F}  smaller than those of powerful radio galaxies that typically exceed $10^{8}$~M$_{\odot}$. 
The detection of GeV \cite{2009ApJ...707L.142A,2012MNRAS.426..317D,2009ApJ...699..976A,2015MNRAS.452..520D} and radio emission \cite{2012A&A...548A.106F,2015A&A...575A..55A,2017A&A...603A.100L} from jetted NLSy1s, challenges the current understanding of relativistic jet formation in which powerful relativistic jets are preferentially found in elliptical galaxies with nuclear black hole masses beyond $10^{8}$~M$_{\odot}$. 
The systematically lower-mass black holes and the accretion with rates close to the Eddington limit ($0.2-0.9$~$L_\mathrm{Edd}$), make this class of source a unique probe of a previously unexplored parameter space. Here we study the optical polarisation variability of 10 selected jetted NLSy1s. All the details are discussed in section~\ref{sec:sample}.  

\section{Previous studies: radio variability}

\begin{figure}[h!] 
\centering
\begin{tabular}{l@{\hskip 0.05cm}l@{\hskip 0.05cm}l@{\hskip 0.05cm}l}
 \includegraphics[trim={0 140 0 0},clip, width=.25\textwidth]{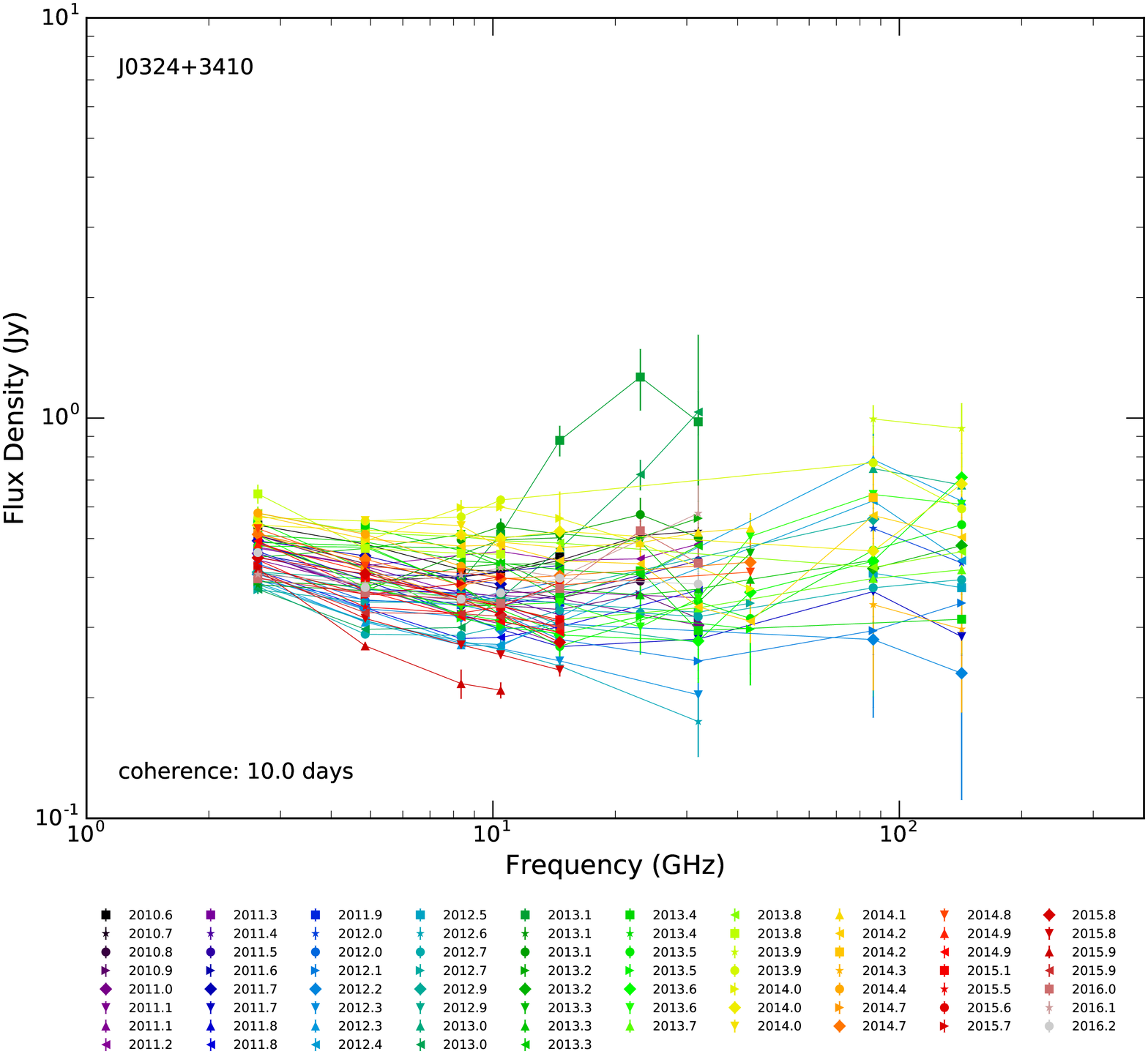} &\includegraphics[trim={60 70 0 0},clip, width=.235\textwidth]{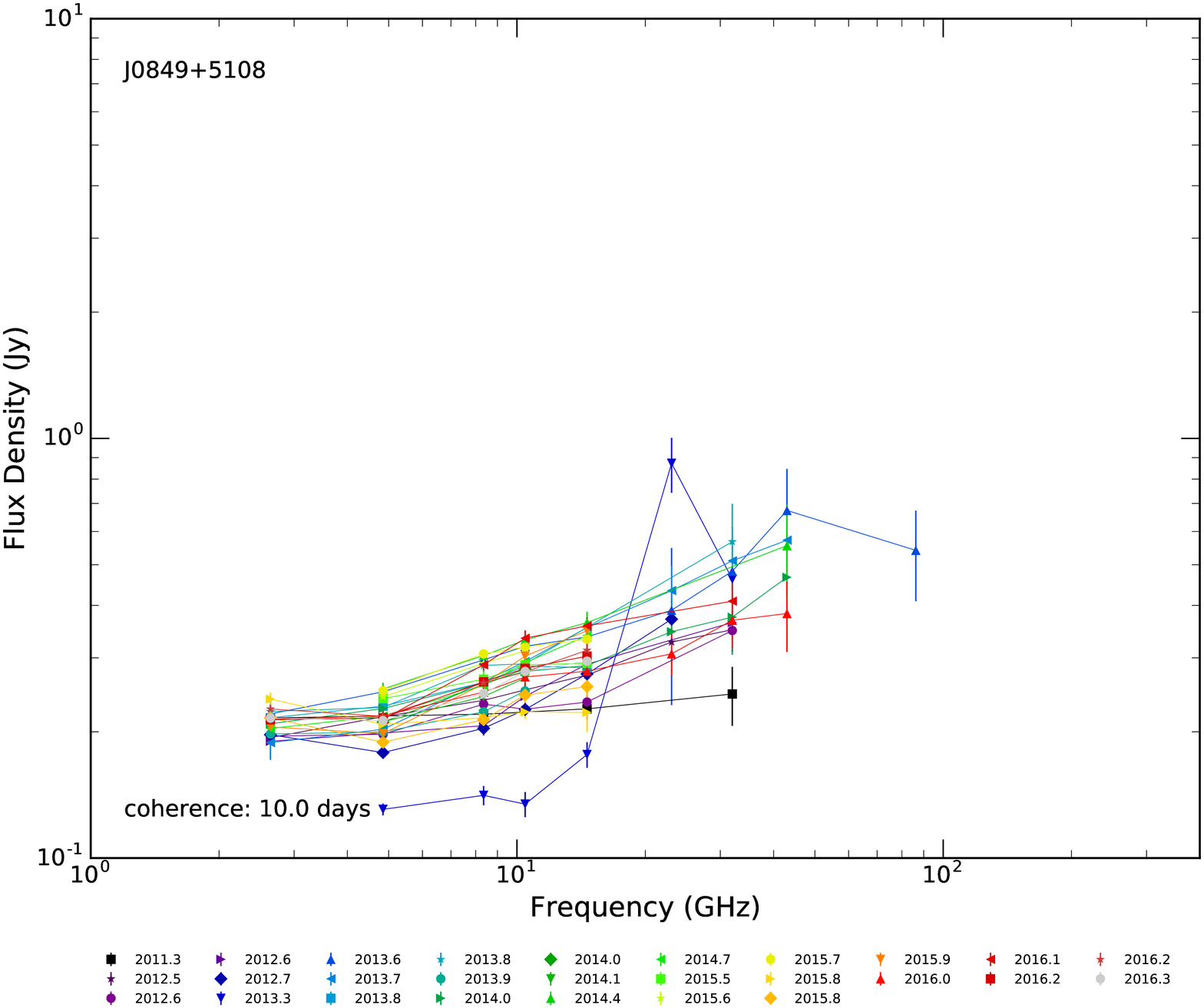}
 &\includegraphics[trim={50 165 0 0},clip, width=.235\textwidth]{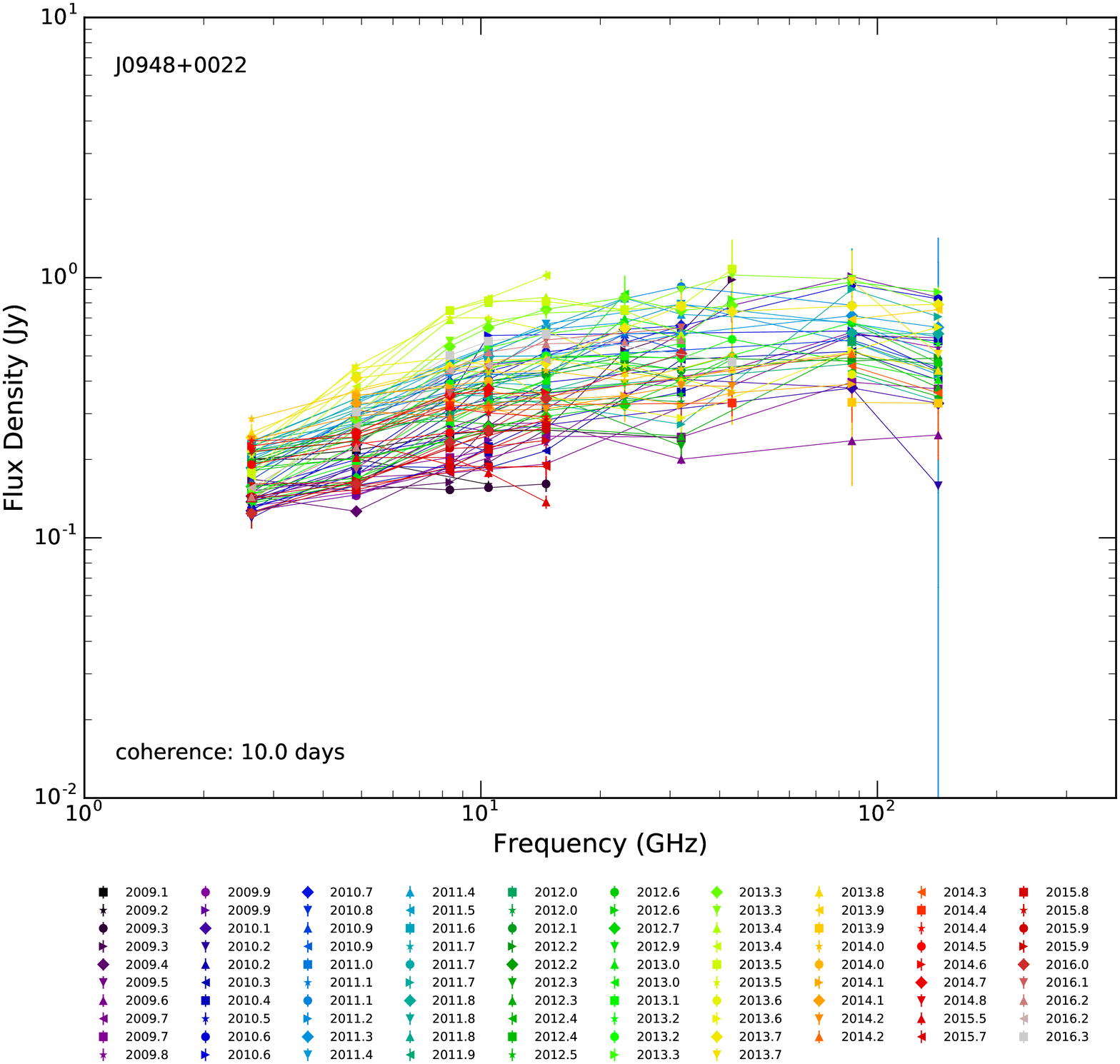} &\includegraphics[trim={50 105 0 0},clip, width=.235\textwidth]{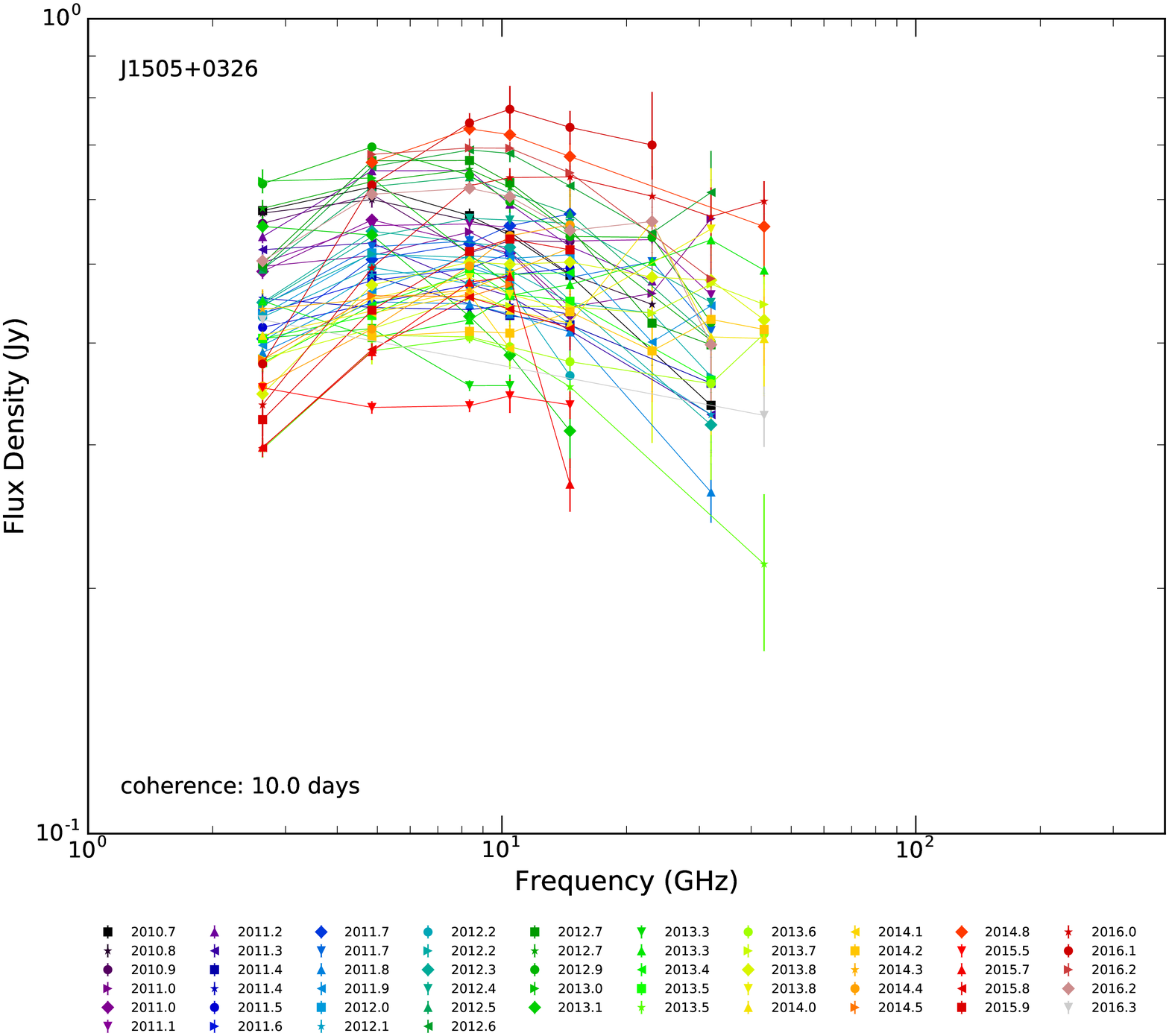}\\
\end{tabular}
\caption{Radio SEDs of the first four NLSy1s that were detected in GeV bands. Till the end of 2015 the nominal cadence is roughly one measurement per month. Afterwards it is almost biweekly. }
\label{fig:specs}
\end{figure}

For the first four NLSy1s detected in GeV energies we initiated a comprehensive multi-frequency monitoring of their radio emission which is discussed in  \cite{2015A&A...575A..55A}. The program -- which is still ongoing -- has been collecting data at 10 frequencies between 2.64~GHz and 142.33~GHz  with initially monthly and later biweekly cadence. All sources show blazar-like behaviour characterised by intense variability and pronounced spectral evolution. From the estimates of the brightness temperatures we infer moderate Doppler factors indicative of rather moderately relativistic jets. Similarly, the jet powers we computed placed them in the class of the least energetic blazars (BL Lac objects). Figure~\ref{fig:specs} illustrates the radio SEDs updated with recently observed data.    

Later we studied the structural dynamics of the NLSy1 1H\,0323$+$342 with 15 GHz MOJAVE data \cite{2016RAA....16..176F}. We found superluminal components exhibiting speeds up to 6.9\,c indicative of a relativistic jet. On the basis of these apparent motions and an estimate of the variability Doppler factor of $\delta_\mathrm{var}\sim 5.2$, we inferred a viewing angle of less than $\sim10^\circ$ confirming the ``aligned jet'' scenario \cite{2009ApJ...707L.142A}.    

\section{Current study: optical polarisation variability}

As a natural next step of our previous studies which indicate the presence of a mildly relativistic jet, we further study the optical polarisation of RL~NLSy1s. Specifically, our aim is to quantify the variability of the optical polarisation and compare it with that of typical blazars. The ultimate aim, however, is to examine whether long rotations of the optical polarisation plane similar to those systematically found in blazars \cite{2016MNRAS.462.1775B} occur also in the case of NLSy1s. And, if so, what is their association with the activity in the GeV energy bands.        

\section{Sample and dataset}
\label{sec:sample}
\begin{figure*}[h] 
\centering
\includegraphics[trim={0 0 0 210},clip, width=0.7\textwidth]{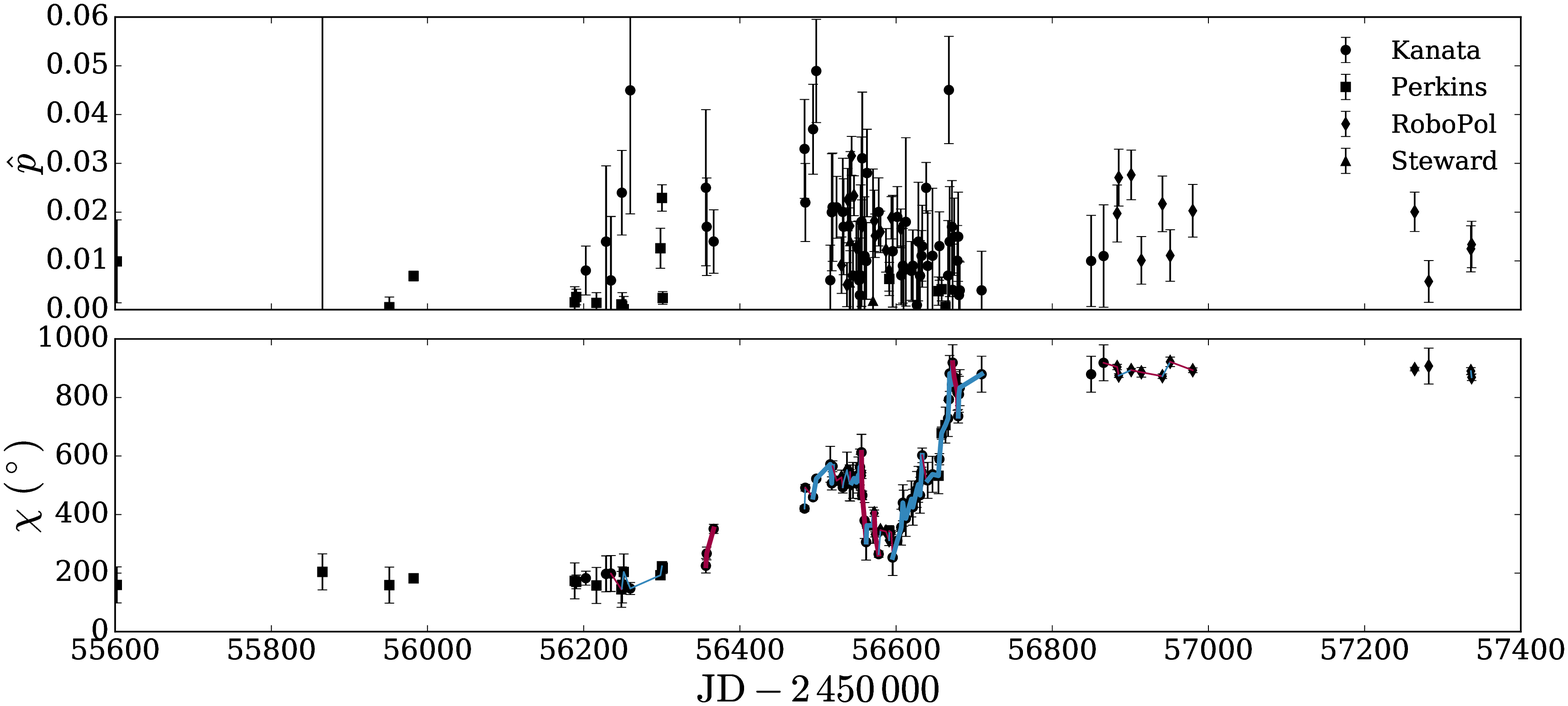}\\ 
\includegraphics[trim={0 0 0 210},clip, width=0.7\textwidth]{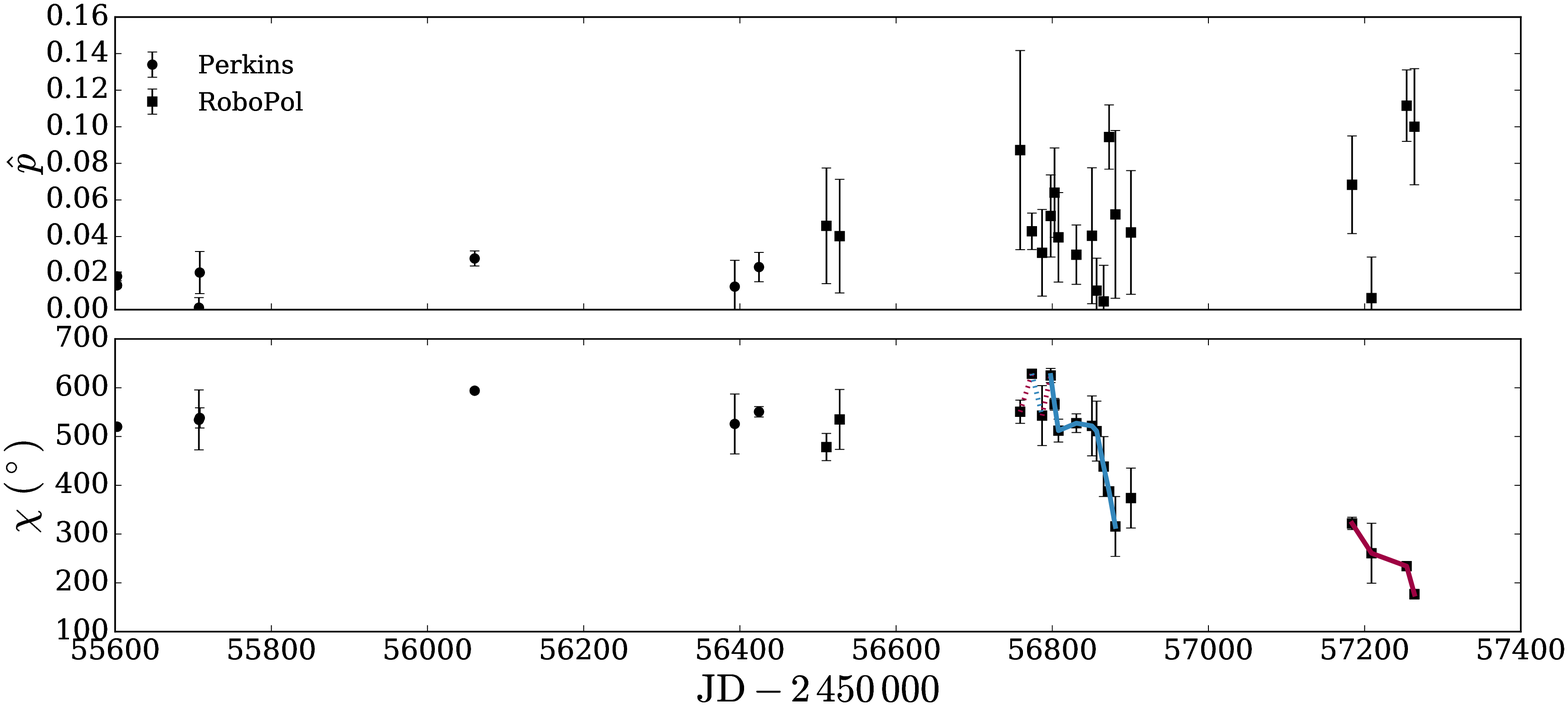} 
\caption{The EVPA ($\chi$) as a function of time. The coloured lines mark periods of significant monotonous -- within the uncertainties -- EVPA evolution. Upper panel is for  J10324$+$3410 and lower panel is for J1505$+$0326.   
Solid lines mark periods of long rotations (i.e. at least three sequential data points and angle larger than $90^\circ$). \textit{Red} and \textit{blue} connecting lines are used alternatively for ease of visualization.}
\label{fig:t_p_evpa_long}
\end{figure*}

We focus on a sample of 10 jetted NLSy1s (table~\ref{tab:sample}) with estimates of the SMBH mass ranging from approximately a few 10$^6$ to a few
  $10^{8}$~M$_{\odot}$, with the majority to lie between $10^{7-8}$~M$_{\odot}$. Five of them are detected by \textit{Fermi} while the remaining five were selected mostly on the basis of their radio-loudness, optical brightness, visibility with the Skinakas telescope and redshift. The latter meant to minimise the host galaxy contribution. Among them, SDSSJ1722$+$5654
  stands out because of its high-amplitude optical variability of $\sim$ 3 mag \cite{2006ApJ...639..710K}. The RoboPol \cite{2014MNRAS.442.1706K} dataset that has been pivoting our research has been augmented with data from the KANATA, Perkins and Steward \cite{2009arXiv0912.3621S} observatories whenever possible. 
The $n\times \pi$ ambiguity innate in the solution of the equation giving the electric vector position angle (EVPA): $\mathrm{EVPA}=0.5\tan^{-1}\left(U/Q\right)$ is resolved by the ``minimum angle variability'' assumption. That is, from the family of solutions we chose the one for which the absolute difference to the previous angle is a minimum. In table~\ref{tab:sample} we also tabulate the median polarisation fraction and its scatter as derived from our study. 
\begin{table}[]
  \caption{List of sources in our sample and their relevant parameters. The last two columns report the median polarisation fraction and its scatter as derived from our study. The radio loudness $R$ is defined as the ratio of the 6~cm flux to the optical flux at 4400~\AA~\cite{Kellermann1989AJ}. Its values are approximate given the high-amplitude variability of most sources in the radio and optical band. }
  \label{tab:sample}  
  \centering                    
  \begin{tabular}{llcrlll} 
    \hline\hline                 
ID            &Survey ID                    &\mc{1}{l}{Redshift}             &\mc{1}{c}{$R$} & Notes &\mc{1}{c}{$\left<\hat{p}\right>$} &\mc{1}{c}{$\sigma_{p}$}  \\
    \hline\\
J0324$+$3410  & 1H\,0323$+$342              &0.062900   & 318 &Fermi   &0.012   &0.016  \\ 
J0849$+$5108  & SBS\,0846$+$513             &0.584701   &1445 &Fermi   &0.100   &0.078   \\ 
J0948$+$0022  & PMN\,J0948$+$0022           &0.585102   & 355 &Fermi   &0.024   &0.028   \\ 
J1305$+$5116  & WISE J130522.75$+$511640.3  &0.787552   & 223 &        &0.040   &0.030    \\
J1505$+$0326  & PKS\,1502$+$036             &0.407882   &1549 &Fermi   &0.010   &0.002    \\ 
J1548$+$3511  & HB89\,1546$+$353            &0.479014   & 692 &        &0.021   &0.024    \\
J1628$+$4007  & RX\,J16290$+$4007           &0.272486   &  29 &        &0.000   &\ldots   \\
J1633$+$4718  & RX\,J1633.3$+$4718          &0.116030   & 166 &        &0.024   &0.004    \\
J1644$+$2619  & FBQS\,J1644$+$2619          &0.145000   & 447 &Fermi   &0.022   &0.015    \\ 
J1722$+$5654  & SDSS\,J172206.02$+$565451.6 &0.425967   & 234 &        &0.000   &\ldots  \\\\    
\hline                                  
  \end{tabular}
\\
\end{table}
Figure~\ref{fig:t_p_evpa_long} shows the EVPA and $p$ curves for the two best-sampled sources, namely J10324$+$3410 (upper panel) and J1505$+$0326 (lower panel). In these plots the reported polarisation fraction has been de-biased (i.e. treating the Rice bias as described in \cite{2014MNRAS.442.1693P}) and the $n\times \pi$ ambiguity has been removed from the EVPA as discussed earlier. 

\section{ EVPA variability}
We study the variability of both polarisation parameters and search for candidate long rotations of the EVPA. Such events are sought among periods of significant EVPA variability (after accounting also for the uncertainties in the EVPA), that (i) consist of at least 3 data points, and (ii) cover polarisation angles more than $90^\circ$. 

\subsection{J1505$+$0326: a possible long rotation of the polarisation plane}

We start with J1505$+$0326 (Fig.~\ref{fig:t_p_evpa_long} lower panel) because it: (i) comprises a study case for our approach for the assessment of the reliability of the detected long rotation, and (ii) because it is the case with the largest probability that the detected apparent rotation is driven by an intrinsic event. From the periods of significant EVPA variability two qualify to long ones and are marked with solid lines in Fig.~\ref{fig:t_p_evpa_long}. For the longest event (lower panel, blue solid line) the polarisation plane rotates by almost $-309.5^\circ$ at a mean rate of $\Delta\chi/\Delta t \approx-3.7$~deg~d$^{-1}$. 

First we discuss the reliability of the observed event. The combination of relatively sparse sampling and the large uncertainties in the EVPA, make the direction of evolution of the EVPA uncertain making the rotation event itself, unreliable. This is the consequence of the fact that the large EVPA uncertainties may allow both $\chi$ and $ \chi + \pi$ to be valid solutions for the angle. We assume that each measurement of the $Q$ and $U$ indeed described the true source polarisation state. By allowing Stokes parameters to oscillate within the range of the observed uncertainties we produce a large number of EVPA curves and compared them to the observed event. We find that the probability for such a curve to be within $1\sigma$ of the observed rotation is around 23\%.        

We then investigate whether the observational noise alone can produce the observed event while the EVPA remains unchanged, i.e. $d\chi_\mathrm{intr}/dt=0$. After $25\times10^4$ iterations we find that the probability of finding a full rotation, is   
\begin{equation}
	\label{eq:J1505_P3}
P\left(\mathrm{full~rotation;}~|\Delta \chi_\mathrm{intr}|\ge 309.5^\circ|~d\chi_\mathrm{intr}/dt=0 \right)=6\times10^{-4}
\end{equation}
fairly lower than the 23\% estimated earlier. From this we conclude that although it is indeed possible that noise alone can cause the observed phenomenology, it is much more unlikely that an intrinsic evolution of the EVPA is driving the observed behaviour.  

On the basis of the assumption that the intrinsic event is obeying a constant rotation rate ($d\chi_\mathrm{intr}/dt=\mathrm{constant}$) we compute the most probable  rate and examine whether the observed rotation is consistent with the predictions of this assumption. We find that for a full rotation over an angle within $1\sigma$ of the observed one the most probable rate is $-3.1$~deg~d$^{-1}$. The distribution of the simulated events shows that indeed the observed angle of rotation is consistent wit the assumption of a constant rotation rate. 

\subsection{J0324$+$3410: the randomness of the EVPA}
\begin{figure}[h!] 
\centering
\begin{tabular}{ll}
 \includegraphics[trim={0 10 0 0},clip, width=.5\textwidth]{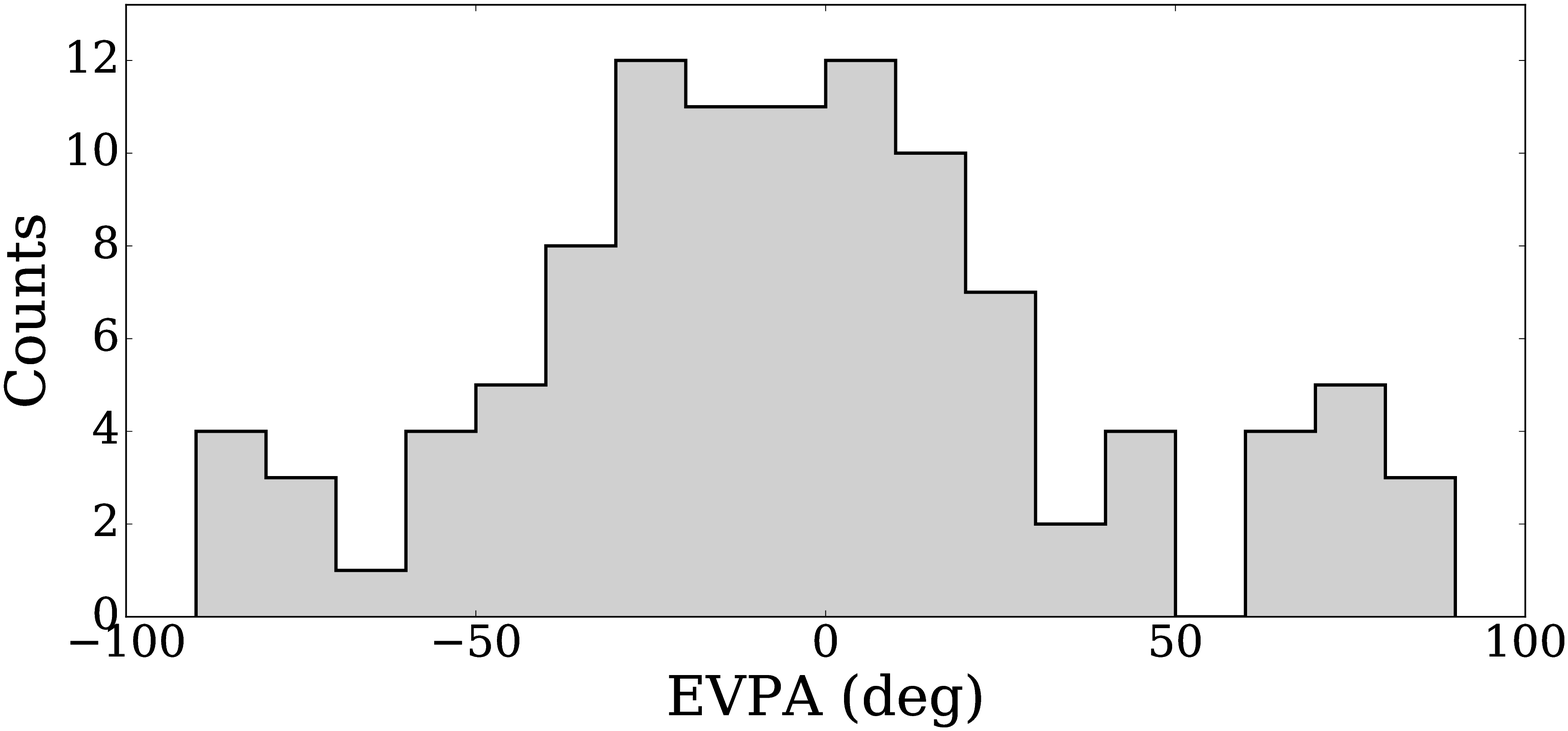} &\includegraphics[trim={0 0 0 40},clip, width=.3\textwidth]{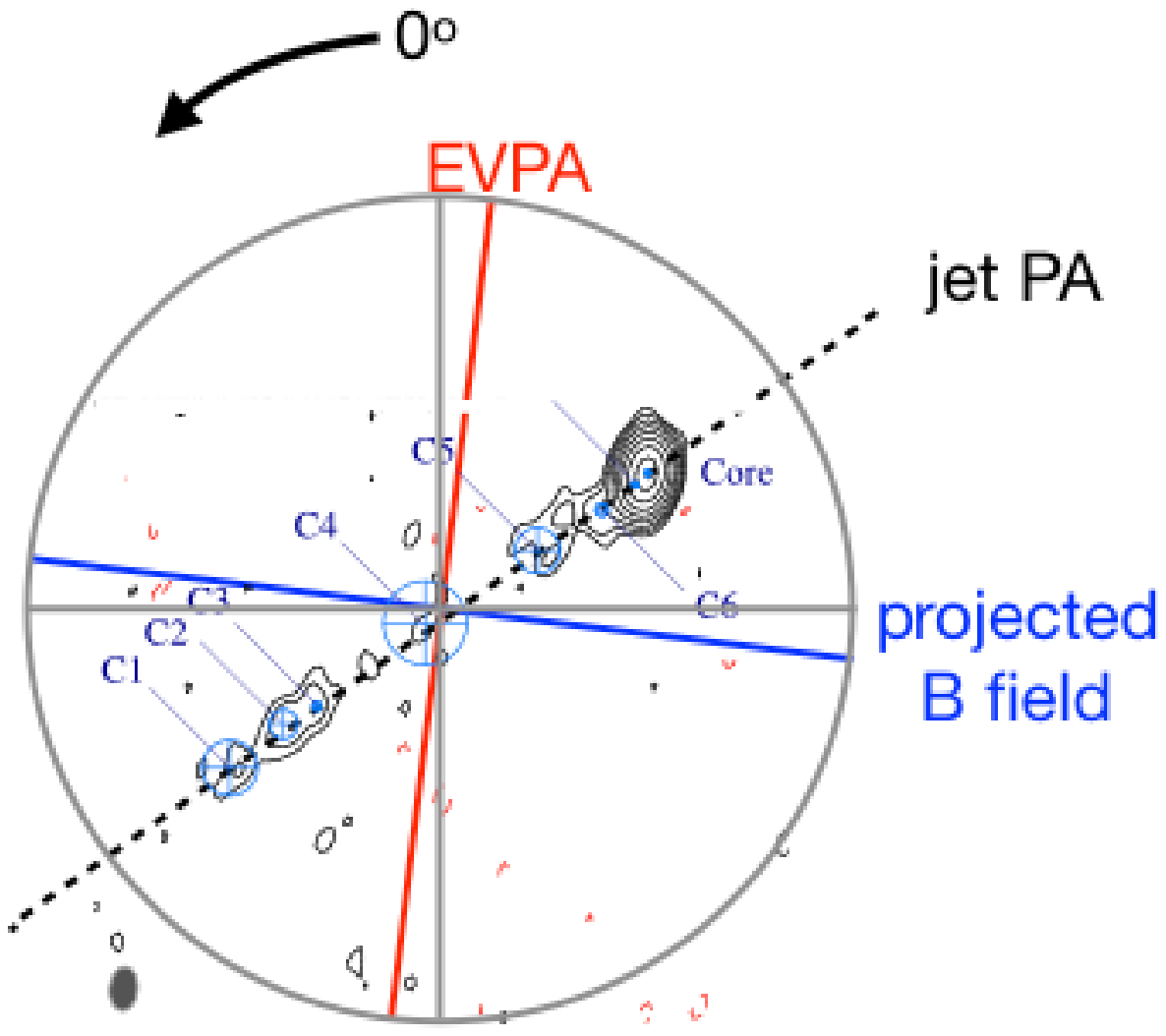}\\
\end{tabular}
\caption{The distribution of the EVPA in the range $\left[-90,+90\right]$ for J10324$+$3410. Right hand panel shows the inferred orientation of the projected magnetic field to the radio jet.}
\label{fig:J0324_evpa_hist}
\end{figure}

Intrigued by the dependence on the synchrotron peak frequency of the randomness of the EVPA \cite{2016MNRAS.463.3365A,2017arXiv171104824A}, we examined the distribution of the EVPA in the best-sampled case in our dataset, J0324$+$3410 (Fig.~\ref{fig:J0324_evpa_hist}). As shown in the left panel of Fig.~\ref{fig:J0324_evpa_hist}
, the EVPA exhibits a preferred angle which can be approximated by the median of $-6.7^\circ$. In \cite{2016RAA....16..176F} we show that the 15 GHz radio jet has a position angle of around $124^\circ$. Assuming then that the transmitting plasma is transparent at optical wavelengths the projected magnetic field (perpendicular to the observed EVPA ) will be at $40.7^\circ$ with the radio jet; this orientation is ambiguous as to whether a poloidal or a toroidal configuration is dominant in that part of the flow (Fig.~\ref{fig:J0324_evpa_hist} right hand panel).            

\section{Conclusions}

After the systematic multi-frequency radio monitoring presented in \cite{2015A&A...575A..55A} and the study of the structural evolution of J0324$+$3410 \cite{2016RAA....16..176F},  we have monitored the $R$-band optical polarisation of a total of 10 jetted NLSy1 galaxies five of which have been detected by \textit{Fermi}-GST. Our findings show that the EVPA undergoes significant variability. In the two best-sampled cases we find candidate long rotations of the polarisation plane similar to those found in blazars. We assess the reliability of these events and we conclude that (at least in one case) although pure noise can induce the observed behaviour, it is much more likely that an intrinsic evolution of the EVPA is taking place. Denser sampling will be necessary for proving the case. Nevertheless, this is the first report of such candidate events in this class of sources; yet another indication for the operation of a blazar-like jet. In the case of  J0324$+$3410 we show that EVPA shows a preferred orientation which corresponds to a projected magnetic field of around $40^\circ$ to the radio jet. This orientation is ambiguous with regards to whether a toroidal or a poloidal field dominates at those scales.

\section*{Acknowledgements}
This conference has been organized with the support of the Department of Physics and Astronomy ``Galileo Galilei'', the University of Padova, the National Institute of Astrophysics INAF, the Padova Planetarium, and the RadioNet consortium. RadioNet has received funding from the European Union's Horizon 2020 research and innovation programme under grant agreement No~730562. 
The RoboPol project is a collaboration between Caltech in the USA, MPIfR in Germany, Torun Centre for Astronomy in Poland, the University of Crete/FORTH in Greece, and IUCAA in India. Data from the Steward Observatory spectropolarimetric monitoring project were used. This program is supported by Fermi Guest Investigator grants NNX08AW56G, NNX09AU10G, NNX12AO93G, and NNX15AU81G. The authors wish to thank the internal MPIfR referee Dr N. MacDonald for the careful reading of the manuscript and the detailed and useful comments.

%

{\footnotesize
\bibliographystyle{JHEP} 
\bibliography{/Users/mangel/work/Literature/MyBIB/References} 

\providecommand{\href}[2]{#2}\begingroup\raggedright\begin{thebibliography}{10}

\bibitem{1985ApJ..297...166}
D.~E. {Osterbrock} and R.~{Pogge}, \emph{{The Spectra of Narrow-Line Seyfert 1
  Galaxies}}, {\emph{\apj} {\bfseries 297} (1985) 166}.

\bibitem{1989ApJ...342..224G}
R.~W. {Goodrich}, \emph{{Spectropolarimetry of 'narrow-line' Seyfert 1
  galaxies}}, \href{https://doi.org/10.1086/167586}{\emph{\apj} {\bfseries 342}
  (1989) 224}.

\bibitem{2006ApJS..166..128Z}
H.~{Zhou}, T.~{Wang}, W.~{Yuan}, H.~{Lu}, X.~{Dong}, J.~{Wang} et~al., \emph{{A
  Comprehensive Study of 2000 Narrow Line Seyfert 1 Galaxies from the Sloan
  Digital Sky Survey. I. The Sample}},
  \href{https://doi.org/10.1086/504869}{\emph{\apjs} {\bfseries 166} (2006)
  128} [\href{https://arxiv.org/abs/arXiv:astro-ph/0603759}{{\ttfamily
  arXiv:astro-ph/0603759}}].

\bibitem{2006AJ....132..531K}
S.~{Komossa}, W.~{Voges}, D.~{Xu}, S.~{Mathur}, H.-M. {Adorf}, G.~{Lemson}
  et~al., \emph{{Radio-loud Narrow-Line Type 1 Quasars}},
  \href{https://doi.org/10.1086/505043}{\emph{\aj} {\bfseries 132} (2006) 531}
  [\href{https://arxiv.org/abs/astro-ph/0603680}{{\ttfamily
  astro-ph/0603680}}].

\bibitem{2008ApJ...685..801Y}
W.~{Yuan}, H.~Y. {Zhou}, S.~{Komossa}, X.~B. {Dong}, T.~G. {Wang}, H.~L. {Lu}
  et~al., \emph{{A Population of Radio-Loud Narrow-Line Seyfert 1 Galaxies with
  Blazar-Like Properties?}}, \href{https://doi.org/10.1086/591046}{\emph{\apj}
  {\bfseries 685} (2008) 801}
  [\href{https://arxiv.org/abs/0806.3755}{{\ttfamily 0806.3755}}].

\bibitem{2012AJ....143...83X}
D.~{Xu}, S.~{Komossa}, H.~{Zhou}, H.~{Lu}, C.~{Li}, D.~{Grupe} et~al.,
  \emph{{Correlation Analysis of a Large Sample of Narrow-line Seyfert 1
  Galaxies: Linking Central Engine and Host Properties}},
  \href{https://doi.org/10.1088/0004-6256/143/4/83}{\emph{\aj} {\bfseries 143}
  (2012) 83} [\href{https://arxiv.org/abs/1201.2810}{{\ttfamily 1201.2810}}].

\bibitem{2015A&A...575A..13F}
L.~{Foschini}, M.~{Berton}, A.~{Caccianiga}, S.~{Ciroi}, V.~{Cracco}, B.~M.
  {Peterson} et~al., \emph{{Properties of flat-spectrum radio-loud narrow-line
  Seyfert 1 galaxies}},
  \href{https://doi.org/10.1051/0004-6361/201424972}{\emph{\aap} {\bfseries
  575} (2015) A13} [\href{https://arxiv.org/abs/1409.3716}{{\ttfamily
  1409.3716}}].

\bibitem{2009ApJ...707L.142A}
A.~A. {Abdo}, M.~{Ackermann}, M.~{Ajello}, L.~{Baldini}, J.~{Ballet},
  G.~{Barbiellini} et~al., \emph{{Radio-Loud Narrow-Line Seyfert 1 as a New
  Class of Gamma-Ray Active Galactic Nuclei}},
  \href{https://doi.org/10.1088/0004-637X/707/2/L142}{\emph{\apjl} {\bfseries
  707} (2009) L142} [\href{https://arxiv.org/abs/0911.3485}{{\ttfamily
  0911.3485}}].

\bibitem{2012MNRAS.426..317D}
F.~{D'Ammando}, M.~{Orienti}, J.~{Finke}, C.~M. {Raiteri}, E.~{Angelakis},
  L.~{Fuhrmann} et~al., \emph{{SBS 0846+513: a new {$\gamma$}-ray-emitting
  narrow-line Seyfert 1 galaxy}},
  \href{https://doi.org/10.1111/j.1365-2966.2012.21707.x}{\emph{\mnras}
  {\bfseries 426} (2012) 317}
  [\href{https://arxiv.org/abs/1207.3092}{{\ttfamily 1207.3092}}].

\bibitem{2009ApJ...699..976A}
A.~A. {Abdo}, M.~{Ackermann}, M.~{Ajello}, M.~{Axelsson}, L.~{Baldini},
  J.~{Ballet} et~al., \emph{{Fermi/Large Area Telescope Discovery of Gamma-Ray
  Emission from a Relativistic Jet in the Narrow-Line Quasar PMN J0948+0022}},
  \href{https://doi.org/10.1088/0004-637X/699/2/976}{\emph{\apj} {\bfseries
  699} (2009) 976} [\href{https://arxiv.org/abs/0905.4558}{{\ttfamily
  0905.4558}}].

\bibitem{2015MNRAS.452..520D}
F.~{D'Ammando}, M.~{Orienti}, J.~{Larsson} and M.~{Giroletti}, \emph{{The first
  {$\gamma$}-ray detection of the narrow-line Seyfert 1 FBQS J1644+2619}},
  \href{https://doi.org/10.1093/mnras/stv1278}{\emph{\mnras} {\bfseries 452}
  (2015) 520} [\href{https://arxiv.org/abs/1503.08226}{{\ttfamily
  1503.08226}}].

\bibitem{2012A&A...548A.106F}
L.~{Foschini}, E.~{Angelakis}, L.~{Fuhrmann}, G.~{Ghisellini}, T.~{Hovatta},
  A.~{Lahteenmaki} et~al., \emph{{Radio-to-{$\gamma$}-ray monitoring of the
  narrow-line Seyfert 1 galaxy PMN J0948 + 0022 from 2008 to 2011}},
  \href{https://doi.org/10.1051/0004-6361/201220225}{\emph{\aap} {\bfseries
  548} (2012) A106} [\href{https://arxiv.org/abs/1209.5867}{{\ttfamily
  1209.5867}}].

\bibitem{2015A&A...575A..55A}
E.~{Angelakis}, L.~{Fuhrmann}, N.~{Marchili}, L.~{Foschini}, I.~{Myserlis},
  V.~{Karamanavis} et~al., \emph{{Radio jet emission from GeV-emitting
  narrow-line Seyfert 1 galaxies}},
  \href{https://doi.org/10.1051/0004-6361/201425081}{\emph{\aap} {\bfseries
  575} (2015) A55} [\href{https://arxiv.org/abs/1501.02158}{{\ttfamily
  1501.02158}}].

\bibitem{2017A&A...603A.100L}
A.~{L{\"a}hteenm{\"a}ki}, E.~{J{\"a}rvel{\"a}}, T.~{Hovatta}, M.~{Tornikoski},
  D.~L. {Harrison}, M.~{L{\'o}pez-Caniego} et~al., \emph{{37 GHz observations
  of narrow-line Seyfert 1 galaxies}},
  \href{https://doi.org/10.1051/0004-6361/201630257}{\emph{\aap} {\bfseries
  603} (2017) A100} [\href{https://arxiv.org/abs/1703.10365}{{\ttfamily
  1703.10365}}].

\bibitem{2016RAA....16..176F}
L.~{Fuhrmann}, V.~{Karamanavis}, S.~{Komossa}, E.~{Angelakis}, T.~P.
  {Krichbaum}, R.~{Schulz} et~al., \emph{{Inner jet kinematics and the viewing
  angle towards the {$\gamma$}-ray narrow-line Seyfert 1 galaxy 1H 0323+342}},
  \href{https://doi.org/10.1088/1674-4527/16/11/176}{\emph{Research in
  Astronomy and Astrophysics} {\bfseries 16} (2016) 176}
  [\href{https://arxiv.org/abs/1608.03232}{{\ttfamily 1608.03232}}].

\bibitem{2016MNRAS.462.1775B}
D.~{Blinov}, V.~{Pavlidou}, I.~{Papadakis}, S.~{Kiehlmann}, I.~{Liodakis},
  G.~V. {Panopoulou} et~al., \emph{{RoboPol: do optical polarization rotations
  occur in all blazars?}},
  \href{https://doi.org/10.1093/mnras/stw1732}{\emph{\mnras} {\bfseries 462}
  (2016) 1775} [\href{https://arxiv.org/abs/1607.04292}{{\ttfamily
  1607.04292}}].

\bibitem{2006ApJ...639..710K}
S.~{Komossa}, W.~{Voges}, H.-M. {Adorf}, D.~{Xu}, S.~{Mathur} and S.~F.
  {Anderson}, \emph{{The Radio-Loud Narrow-Line Quasar SDSS
  J172206.03+565451.6}}, \href{https://doi.org/10.1086/499764}{\emph{\apj}
  {\bfseries 639} (2006) 710}
  [\href{https://arxiv.org/abs/astro-ph/0511496}{{\ttfamily
  astro-ph/0511496}}].

\bibitem{2014MNRAS.442.1706K}
O.~G. {King}, D.~{Blinov}, A.~N. {Ramaprakash}, I.~{Myserlis}, E.~{Angelakis},
  M.~{Balokovi{\'c}} et~al., \emph{{The RoboPol pipeline and control system}},
  \href{https://doi.org/10.1093/mnras/stu176}{\emph{\mnras} {\bfseries 442}
  (2014) 1706} [\href{https://arxiv.org/abs/1310.7555}{{\ttfamily 1310.7555}}].

\bibitem{2009arXiv0912.3621S}
P.~S. {Smith}, E.~{Montiel}, S.~{Rightley}, J.~{Turner}, G.~D. {Schmidt} and
  B.~T. {Jannuzi}, \emph{{Coordinated Fermi/Optical Monitoring of Blazars and
  the Great 2009 September Gamma-ray Flare of 3C 454.3}}, {\emph{ArXiv
  e-prints} (2009) } [\href{https://arxiv.org/abs/0912.3621}{{\ttfamily
  0912.3621}}].

\bibitem{Kellermann1989AJ}
K.~I. {Kellermann}, R.~{Sramek}, M.~{Schmidt}, D.~B. {Shaffer} and R.~{Green},
  \emph{{VLA observations of objects in the Palomar Bright Quasar Survey}},
  \href{https://doi.org/10.1086/115207}{\emph{\aj} {\bfseries 98} (1989) 1195}.

\bibitem{2014MNRAS.442.1693P}
V.~{Pavlidou}, E.~{Angelakis}, I.~{Myserlis}, D.~{Blinov}, O.~G. {King},
  I.~{Papadakis} et~al., \emph{{The RoboPol optical polarization survey of
  gamma-ray-loud blazars}},
  \href{https://doi.org/10.1093/mnras/stu904}{\emph{\mnras} {\bfseries 442}
  (2014) 1693} [\href{https://arxiv.org/abs/1311.3304}{{\ttfamily 1311.3304}}].

\bibitem{2016MNRAS.463.3365A}
E.~{Angelakis}, T.~{Hovatta}, D.~{Blinov}, V.~{Pavlidou}, S.~{Kiehlmann},
  I.~{Myserlis} et~al., \emph{{RoboPol: the optical polarization of
  gamma-ray-loud and gamma-ray-quiet blazars}},
  \href{https://doi.org/10.1093/mnras/stw2217}{\emph{\mnras} {\bfseries 463}
  (2016) 3365} [\href{https://arxiv.org/abs/1609.00640}{{\ttfamily
  1609.00640}}].

\bibitem{2017arXiv171104824A}
E.~{Angelakis}, D.~{Blinov}, M.~{B{\"o}ttcher}, T.~{Hovatta}, S.~{Kiehlmann},
  I.~{Myserlis} et~al., \emph{{The dependence of optical polarisation of
  blazars on the synchrotron component peak frequency}}, {\emph{ArXiv e-prints}
  (2017) } [\href{https://arxiv.org/abs/1711.04824}{{\ttfamily 1711.04824}}].

\end{thebibliography}\endgroup
}

\end{document}